# A density functional theory for symmetric radical cations from bonding to dissociation

**Ester Livshits and Roi Baer***

*Institute of Chemistry and the Fritz Haber Center for Molecular Dynamics, the Hebrew University of Jerusalem, Jerusalem 91904 Israel.*

It is known for quite some time that approximate density functional (ADF) theories fail disastrously when describing the dissociative symmetric radical cations $R_2^+$. Considering this dissociation limit, previous work has shown that Hartree-Fock (HF) theory favors the $R^{+1}$-$R^0$ charge distribution while DF approximations favor the $R^{+0.5}$-$R^{+0.5}$. Yet, general quantum mechanical principles indicate that both these (as well as all intermediate) average charge distributions are asymptotically energy degenerate. Thus HF and ADF theories mistakenly break the symmetry but in a contradicting way. In this letter we show how to construct system-dependent long-range corrected (LC) density functionals that can successfully treat this class of molecules, avoiding the spurious symmetry breaking. Examples and comparisons to experimental data is given for R=H, He and Ne and it is shown that the new LC theory improves considerably the theoretical description of the $R_2^+$ bond properties, the long range form of the asymptotic potential curve as well as the atomic polarizability. The broader impact of this finding is discussed as well and it is argued that the widespread semi-empirical approach which advocates treating the LC parameter as a system-independent parameter is in fact inappropriate under general circumstances.

At large inter-nuclear distance $r$ the electronic ground state wavefunction of symmetric radical cations $R_2^+$ is excellently approximated by the superposition

$$|\theta\rangle = |10\rangle \cos\theta + |01\rangle \sin\theta, \qquad (1)$$

where $\theta$ is an arbitrary angle and $|01\rangle$ ($|10\rangle$) is the exact wavefunction having positive charge on the right (left) R fragment. In the asymptotic $r \to \infty$ limit, the *average* charge of the left R fragment is $Q_L(\theta) = e\cos^2\theta$ while that on the right is $Q_R(\theta) = e\sin^2\theta$. Because $|10\rangle$ and $|01\rangle$ are energy degenerate, the ground state energy $E(r,\theta)$ is almost <u>independent</u> of $\theta$ at large $r$ (at $r \to \infty$ this is exact). How well this degeneracy is preserved in various approximations is a central issue in this paper. The energy difference at some large inter-nuclear distance $r_0 = 20$Å,

$$\Delta E(r_0) = E\left(r_0, \frac{\pi}{4}\right) - E(r_0, 0) \qquad (2)$$

will serve as a measure for this. Note that $\theta = \frac{\pi}{4}$ is the symmetric charge distribution while $\theta = 0$ (or $\frac{\pi}{2}$) is a localized (integer) average charge distribution. An additional exact property of the potential is its functional form at large $r$, which must approach an atom-ion interaction potential[1]:

$$E(r,\theta) \to 2E_R - IP_R - \frac{2\alpha_R}{r^4} + O(r^{-6}), \qquad (3)$$

where, $E_R$ is the electronic energy of the atom $R$, $IP_R$ is its ionization potential and $\alpha_R$ its polarizability (atomic units are used throughout).

The above exact properties, derived from compelling physical considerations, make the class of symmetric cation systems an interesting and generic benchmark for DFT. Surprisingly, local approximations to DFT have been found to grossly defy them exhibiting a repulsive asymptotic Born-Oppenheimer (BO) force $F(r) = -E'(r) \propto r^{-2}$ at large distances of the attractive force deduced from Eq. (3).[2] It was further found that the approximate density functionals (ADF) break the $\theta$ degeneracy, stabilizing the delocalized charge distribution $\Delta E_{ADF} > 0$. Equally interesting was the fact that Hartree-Fock theory also broke symmetry but in an opposite way, stabilizing the delocalized charge density: $\Delta E_{HF} < 0$. As shown below, this spurious behavior often spoils the accuracy of the potential also at small $r$.

It was established in previous studies that the spurious asymptotic repulsion is a result of self-repulsion in conventional DF approximations[2-8]. It is therefore appealing to determine how fares the long-range corrected (LC) hybrid functional,[9-13] with these systems. In the LC approach one separates the electron repulsion potential appearing in the exchange terms into short- and long-range parts: $r^{-1} = r^{-1}erf(\gamma r) + r^{-1}erfc(\gamma r)$. The short range term is represented by a local potential while the long range part is treated via an "explicit" or "exact" exchange term. One critical issue in this scheme is the determination of $\gamma$ and any additional parameter inserted into the scheme. If one assumes that $\gamma$ is system independent one can use a molecular training set for optimizing its value. Such semi-empirical approaches were shown to achieve impressive results for a limited class of systems[11,14-18]. However, assuming $\gamma$ is system independent is most likely only an approximation. Indeed, in ref. [12] a rigorous theory for $\gamma$ was developed, based on the adiabatic connection theorem. The value of $\gamma(n)$ was subsequently computed for the homogeneous electron gas[18,19] using high accuracy Monte Carlo results,[20] finding strong density dependence. Furthermore, an ab-initio method for deter-

---

mining system-specific $\gamma$ showed good predictions of ionization potentials [18]

One of the goals of the present letter is to show that a <u>system-dependent</u> parameter $\gamma$ is in fact necessary for treating the symmetric radical cation systems. We also provide a simple ab initio approach for determining its proper value. While our approach can be used with any existing LC functional, we focus here on a specific functional developed in ref. [18] which we label here as BNL. All results below were calculated using the Q-CHEM program version 3.1[21] and the basis set was cc-pVTZ (except where indicated).

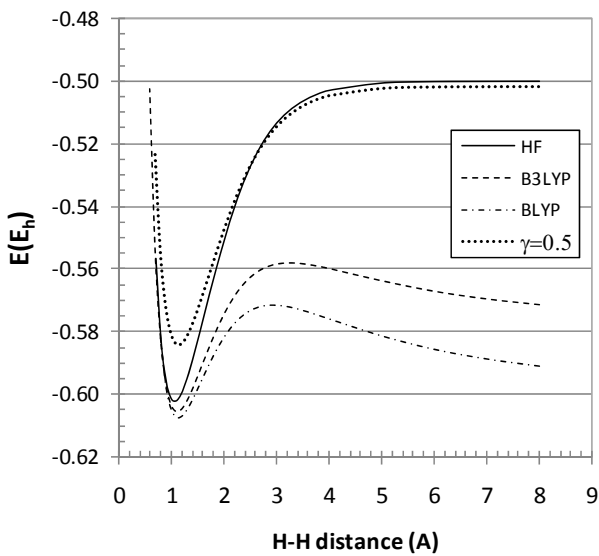

Figure 1: The potential curves of $H_2^+$ in a cc-pVTZ basis using different theories. The "HF" theory is the exact variational solution in this case since this is a 1-electron system.

Let us first consider the simplest $R_2^+$ system, namely $H_2^+$, for which HF results are essentially exact within the basis set. In Figure 1 we show the potential energy curves $E(r)$ of $H_2^+$ calculated using HF, BLYP, B3LYP, and LC BNL ($\gamma = 0.5$) functionals. In the face of the exact HF curve, all DFT curves look faulty. The BLYP and B3LYP expose their unwieldy dissociation pattern and the BNL functional based curve, while being qualitatively correct is quantitatively disappointing with its underestimated well depth.

So, let us now discuss how to improve the functional performance for these systems. We found it is essential to concentrate on the fixing the spurious symmetry breaking. For any value of $\gamma$, we calculate $\Delta E$ of Eq. (2), denoting it $\Delta E_\gamma$. In Figure 2 we show the results for $R_2^+$ with R=H, He and Ne. It is seen that only for a particular value of $\gamma$ does the exact condition $\Delta E_\gamma = 0$ materialize (we denote this as the *proper value* $\gamma^*$). For R = H $\gamma^* \to \infty$ and for R = He and Ne the proper values are $\gamma^* \approx 1.4 a_0^{-1}$ and $\gamma^* \approx 0.93 a_0^{-1}$ respectively. BNL with the proper LC parameter $\gamma^*$ will be denoted henceforth BNL*.

We now show that the proper functional BNL* improves considerably the asymptotic properties of the PES and at the same time gives a good description of the bonding characteristics. In Table 1 we show the essential results. Consider first the enthalpy. We give two ways for estimating it. First, using atomization energies as $E_{at} - (E(r_{eq}) - \hbar\omega/2)$ and second using the asymptotic form of the potential curve $E(\infty) - (E(r_{eq}) - \hbar\omega/2)$. HF theory is exact for R=H, not so good for R=He where the error exceeds 20% and completely wrong for R = Ne where the predicted enthalpy is less than 10% of its experimental value. When BLYP and B3LYP are used to estimate enthalpy via the first method they break up completely when the asymptote method is used. When the atomization energy is used they predict too large enthalpies (up to a factor of 2 for R = Ne). BNL does not break up as BLYP and B3LYP because it has long range self repulsion removed by the two methods for calculating enthalpies differ by a large amount, a manifestation of the spurious symmetry breaking. The proper functional BNL* seems to be very robust and yields in both methods almost the same enthalpy estimates deviating from experimental values typically by less than 10%. Next, consider the equilibrium bond length and the vibrational frequency. Here too, the conventional DFT approximations have larger than usual errors, while HF and the two LC approaches are pretty robust. Finally, the correct asymptotic form of the PES, namely Eq. (3) can be checked by computing the ratio $\alpha_R/\alpha_{eff}$, where $\alpha_R$ is the polarizability of $R$ calculated separately in the same basis and:

$$\alpha_{eff} = \lim_{r \to \infty} E'(r) r^5 / 2 \qquad (4)$$

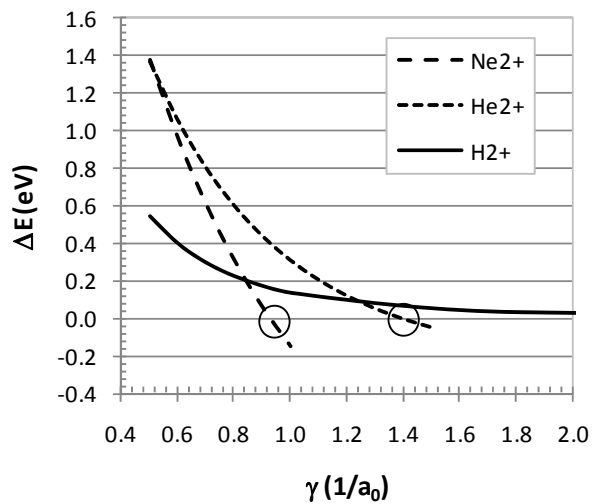

Figure 2: The dependence of the localization-delocalization energy difference $\Delta E$ on the value of the parameter $\gamma$ in the LC functional.

From Eq. (3) it is evident that $\alpha_R/\alpha_{eff}$ should approach 1 as the inter-nuclear distance is increased. Thus, rhe proximity of the ratio $\alpha_R/\alpha_{eff}$ to its ideal value of 1, is a quality indicator for the asymptotic form of the potential. For BNL ($\gamma = 0.5$) we see in Table 1 that the ratio is 0.6 for $H_2^+$ while for the other species the potential does not show a converged value for $\alpha_{eff}$, indicating that it exhibits the wrong asymptotic behavior. The approximate DFTs get the asymptotic all wrong and so the ratio is meaningless. This is specified by "NA" in the table. The Hartree-Fock and BNL* methods have ratios close to 1, indicating proper asymptotic behavior.

Finally, we have checked an atomic property namely the polarizability against experiments. Here near basis-set convergence is needed, so we used the doubly augmented cc-pVTZ basis set (daug-cc-pVTZ). It is seen that BLYP, B3LYP and BNL all tend to overestimate the polarizability while HF underestimates it. The performance of BNL* is exceptionally good with deviations smaller than 3%.

Table 1: Data for $R_2^+$, R=H, He and Ne calculated in the cc-pVTZ basis. All energies in kcal/mole. $r_{eq}$ is the inter-nuclear distance at which the $E(r)$ is minimal. $\omega = \sqrt{\mu E''(r_{eq})}$ is the harmonic frequency and "Enthalpy" is calculated first as $E_{at} - E(r_{eq}) + \frac{1}{2}\hbar\omega$ where $E_{at} = E_R + E_{R^+}$ is the sum of energies of the $R$ atom and the $R^+$ ion and then using the asymptote, as $E(\infty) - E(r_{eq}) + \frac{1}{2}\hbar\omega$. $\alpha_R$ is the calculated polarizability of the atom $R$ and $\alpha_{eff}$ is defined in (4). Finally $\alpha_R^{daug}$ is the polarizability computed with a large basis-set.

| Property | R | BLYP | B3LYP | HF | BNL | BNL* | Exp.[22,23] |
|---|---|---|---|---|---|---|---|
| Enthalpy by atomization | H | 66 | 65 | 60.9 | 60.9 | 60.9 | 61 |
| | He | 82 | 75 | 43 | 74 | 59 | 55 |
| | Ne | 75 | 60 | 2 | 59 | 34 | 32 |
| Enthalpy by asymptote | H | NA | NA | 60.9 | 50 | 60.9 | 61 |
| | He | NA | NA | 43 | 42 | 59 | 55 |
| | Ne | NA | NA | 2 | 27 | 34 | 32 |
| $r_{eq}$ (Å) | H | 1.1 | 1.1 | 1.06 | 1.2 | 1.06 | 1.05 |
| | He | 1.2 | 1.1 | 1.075 | 1.2 | 1.078 | 1.080 |
| | Ne | 1.9 | 1.9 | 1.7 | 1.760 | 1.72 | 1.765 |
| $\hbar\omega/2$ | H | 2.7 | 2.9 | 3.3 | 2.9 | 3.3 | 3.32 |
| | He | 1.7 | 2.0 | 2.5 | 2.1 | 2.5 | 2.42 |
| | Ne | 0.5 | 0.6 | 0.9 | 0.726 | 0.8 | 0.729 |
| $\alpha_R/\alpha_{eff}$ | H | NA | NA | 1 | 0.6 | 1 | 1 |
| | He | NA | NA | 0.98 | NA | 0.98 | 1 |
| | Ne | NA | NA | 1.01 | NA | 1.02 | 1 |
| $\alpha_R^{daug}(a_0^3)$ | H | 5.3 | 5.6 | 4.51 | 5.8 | 4.51 | 4.50 |
| | He | 1.6 | 1.5 | 1.34 | 1.8 | 1.41 | 1.38 |
| | Ne | 3.1 | 2.9 | 2.4 | 3.2 | 2.70 | 2.66 |

In summary, we have shown that by optimizing $\gamma$ so that the asymptotic degeneracy is respected we obtain a functional with improved properties: it automatically adheres to the asymptotic form (Eq. (3)) and it describes the basic bond properties reasonably well and gives very good atomic polarizability. Overall, the LC functional with optimized $\gamma$ is considerably more robust and physically appealing than any of its counterparts considered here.

The broader impact of these findings is that the $\gamma$ parameter must not be considered a system-independent quantity. It seems that for universality to be achieved, $\gamma$ must be system specific and techniques for its ab initio determination from the calculation itself, as done here, need to be developed. Previous work of ours[18] indicated that having a system-dependent $\gamma$ may impair size consistency. Yet, here we found that by choosing $\gamma$ so as to *impose* this consistency we obtain a good description from bonding to dissociation. This approach can perhaps be made into a general strategy so that before any system $R$ is calculated, the parameter $\gamma$ can be determined by the above procedure for $R_2^+$. This will in effect serve to impose the linear energy dependence on the fractional particle number as determined by Perdew et al.[24] exhibiting both derivative discontinuity and correct long-range behavior (we refer the reader to a recent comprehensive review discussing these two aspects[25]). In future work we will examine whether this or similar approaches have the capability of improve other asymptotic properties such as reaction barriers.

**Acknowledgements** We gratefully acknowledge the Israel Science Foundation founded by the Israel Academy of Sciences and Humanities for supporting this work.

[1] A. J. Stone, *The theory of intermolecular forces*. (Oxford University Press, Oxford, 1996).
[2] R. Merkle, A. Savin, and H. Preuss, J. Chem. Phys. **97** (12), 9216 (1992).
[3] T. Bally and G. N. Sastry, J. Phys. Chem. A **101** (43), 7923 (1997).
[4] M. Lundberg and P. E. M. Siegbahn, J. Chem. Phys. **122** (22) (2005).
[5] J. P. Perdew, A. Ruzsinszky, G. I. Csonka, O. A. Vydrov, G. E. Scuseria, V. N. Staroverov *et al.*, Phys. Rev. A **76** (4), 040501 (2007).
[6] O. A. Vydrov, G. E. Scuseria, and J. P. Perdew, J. Chem. Phys. **126** (15), 154109 (2007).
[7] A. Ruzsinszky, J. P. Perdew, G. I. Csonka, O. A. Vydrov, and G. E. Scuseria, J. Chem. Phys. **126** (10), 104102 (2007).
[8] J. Grafenstein, E. Kraka, and D. Cremer, Phys. Chem. Chem. Phys. **6** (6), 1096 (2004).
[9] A. Savin, in *Recent Advances in Density Functional Methods Part I*, edited by D. P. Chong (World Scientific, Singapore, 1995), pp. 129.
[10] T. Leininger, H. Stoll, H.-J. Werner, and A. Savin, Chem. Phys. Lett. **275**, 151 (1997).
[11] H. Iikura, T. Tsuneda, T. Yanai, and K. Hirao, J. Chem. Phys. **115** (8), 3540 (2001).
[12] R. Baer and D. Neuhauser, Phys. Rev. Lett. **94**, 043002 (2005).
[13] T. Yanai, D. P. Tew, and N. C. Handy, Chem. Phys. Lett. **393**, 51 (2004).
[14] M. J. G. Peach, T. Helgaker, P. Salek, T. W. Keal, O. B. Lutnaes, D. J. Tozer *et al.*, Phys. Chem. Chem. Phys. **8** (5), 558 (2005).
[15] O. A. Vydrov and G. E. Scuseria, J. Chem. Phys. **125** (23), 234109 (2006).
[16] Y. Zhao and D. G. Truhlar, J. Phys. Chem. A **110** (49), 13126 (2006).
[17] J. D. Chai and M. Head-Gordon, J. Chem. Phys. **128** (8), 084106 (2008).
[18] E. Livshits and R. Baer, Phys. Chem. Chem. Phys. **9** (23), 2932 (2007).
[19] R. Baer, E. Livshits, and D. Neuhauser, Chem. Phys. **329**, 266 (2006).
[20] P. Gori-Giorgi and J. P. Perdew, Phys. Rev. B **66**, 165118 (2002).

[21] Y. Shao, L. F. Molnar, Y. Jung, J. Kussmann, C. Ochsenfeld, S. T. Brown *et al.*, Phys. Chem. Chem. Phys. **8** (27), 3172 (2006).
[22] K. P. Huber and G. Herzberg, *Molecular Spectra and Molecular Structure, Vol 4: Constants of Diatomic Molecules*. (Prentoce-Hall, New York, 1979).
[23] A. Dalgarno, Adv. Phys. **11** (44), 281 (1962).
[24] J. P. Perdew, R. G. Parr, M. Levy, and J. L. Balduz, Phys. Rev. Lett. **49** (23), 1691 (1982).
[25] S. Kummel and L. Kronik, Rev. Mod. Phys. **80** (1), 3 (2008).